\documentstyle{elsart}
\begin{document}

\begin{frontmatter}

\title{Computer algebra in gravity:\\ Programs for (non-)Riemannian
    spacetimes. I}

\author[koeln,leon]{Jos\'e Socorro},
\author[mex]{Alfredo Mac\'{\i}as}, and
\author[koeln]{Friedrich W. Hehl}
\address[koeln]{Institute for Theoretical Physics, University of
  Cologne\\ D-50923 K\"oln, Germany} 
\address[mex]{ Departamento de F\'{\i}sica,
Universidad Aut\'onoma Metropolitana--Iztapalapa,\\
Apartado Postal 55-534, C.P. 09340, M\'exico, D.F., Mexico}
\address[leon]{Permanent address: Instituto
    de Fisica de la Universidad de Guanajuato, Apartado Postal E-143,
    C.P. 37150, Le\'on, Guanajuato, Mexico.}

\begin{abstract}
  Computer algebra programs are presented for application in general
  relativity, in electrodynamics, and in gauge theories of gravity.
  The mathematical formalism used is the calculus of exterior
  differential forms, the computer algebra system applied Hearn's
  Reduce with Schr\"ufer's exterior form package Excalc. As a
  non-trivial example we discuss a {\em metric} of Pleba\'nski \&
  Demia\'nski (of Petrov type D) together with an electromagnetic
  potential and a {\em triplet of post-Riemannian one-forms}. This
  whole geometrical construct represents an exact solution of a
  metric-affine gauge theory of gravity. We describe a sample session
  and verify by computer that this exact solution fulfills the
  appropriate field equations.-- Computer programs are described for
  the irreducible decomposition of (non-Riemannian) curvature,
  torsion, and nonmetricity.  {\em file cpc9.tex, 1998-03-18}
\end{abstract}
\end{frontmatter}

\section{Introduction}

General relativity (GR), that is, Einstein's gravitational theory, is
one of those scientific subjects in which computer algebra methods
were used as soon as they became available. In fact, in some
instances, general relativity was taken as a guinea pig since this
theory is notorious for its lengthy and messy calculations. Just
remember that the Riemann curvature 2-form $R_{\alpha\beta}:=
g_{\beta\gamma}\,R_\alpha{}^\gamma=-R_{\beta\alpha}$, here
$g_{\beta\gamma}$ are the components of the metric, characterizes the
local geometry of four-dimensional spacetime. It has, in these
dimensions, 20 independent components. And Einstein's field equation
has 10 independent components. These numbers alone make it plausible
that it is desirable to give the corresponding calculations to a
computer, see Brans \cite{Brans}. Since they are analytic, numerics
doesn't help, rather symbol manipulation programs or, as they are
called today, computer algebra systems are necessary.

This is even more important for generalizations of Einstein's theory,
like the Einstein--Cartan theory, the Poincar\'e gauge theory, or the
metric-affine theory of gravity.  In these {\em gauge theories of
  gravitation,} see Gronwald \& Hehl \cite{Erice95} for a review,
spacetime carries additional post-Riemannian structures, namely {\em
  torsion} $T^\alpha$ (a vector valued 2-form) with $6\times 4$
independent components, and/or {\em nonmetricity}
$Q_{\alpha\beta}=Q_{\beta\alpha}$ (a symmetric tensor-valued 1-form)
with $4\times 10$ components.  Moreover, in general, the curvature
loses its antisymmetry. Thus, a metric-affine space, in four
dimensions, is described by the $36+24+40$ `field strengths'
$R_{\alpha\beta}$, $T^\alpha$, and $Q_{\alpha\beta}$, respectively.
For some physicist this is too much and considered to be a plethora of
arbitrariness.  This is like the argument against Einstein's theory,
which was taken quite seriously in some quarters, that the $10$
gravitational potentials of Einstein's theory are too much as compared
to only $1$ potential in Newton's gravitational theory. But
consistency and beauty cannot be measured in numbers of components! In
\cite{PRs}, e.g., one can find the arguments which make us believe
that the microstructure of spacetime requires post-Riemannian degrees
of freedom as represented by torsion and nonmetricity.

Let us come back to {\em general relativity}. A readable survey of the
computer algebra systems in use for general relativity has been given
by Hartley \cite{David}. In the same book \cite{Honnef}, see also
\cite{GR14CA}, there is a whole chapter on computer algebra methods as
used in relativity.  Suppose you would like to use such methods in
general relativity or gauge theories of gravity. Considering the ample
computer resources one has available today, it seems reasonable to us
to turn to a {\em general purpose system} like Macsyma, Maple,
Mathematica, and/or Reduce and, on top of one of these, to use
specialized packages for general relativity/differential geometry, see
\cite{David}. For getting familiar with the corresponding general
relativity packages, we can recommend the respective articles or books
of McLenaghan \cite{McL} for Maple, of Parker \& Christensen
\cite{Parker} and Soleng \cite{Harald} for Mathematica, and of McCrea
\cite{Dermott} for Reduce. Some specific relativistic applications
have been discussed, e.g., in Maple, see \cite{Pollney}, in
Mathematica, see \cite{Klioner,Tsantilis}, and in Reduce, see
\cite{Wolf}.

The general relativity computer algebra packages can, as a rule, also
be used for non-Riemannian spacetimes and the corresponding
gravitational gauge theories. Usually the inclusion of torsion and
nonmetricity is not difficult to program. However, there is not too
much literature available on that subject. For torsion theories there
exist, amongst others, the papers of {\AA}man et al.
\cite{TCLASSI,Aman}, McCrea \cite{Dermott}, Schr\"ufer et al.
\cite{EXCALC}, Tertychniy and Obukhova \cite{Irina}, and Zhytnikov
\cite{GRG}. The packages $GRG$ \cite{GRG} and $GRG_{\rm EC}$
\cite{Irina} (EC stands for Einstein-Cartan theory) are Reduce based
systems.

There is even less literature available for {\em metric-affine spaces}
and the corresponding gauge theories. Such a spacetime carries,
besides the coframe (of 1-forms) $\vartheta^\alpha=e_i{}^\alpha\,
dx^i$, the {\em metric} $g=g_{\alpha\beta}\, \vartheta^\alpha \otimes
\vartheta^\beta=g_{ij}\,dx^i\otimes dx^j$ and the linear {\em
  connection} 1-form $\Gamma_\alpha{}^\beta=\Gamma_{\gamma
  \alpha}{}^\beta\, \vartheta^\gamma=\Gamma_{i\alpha}{}^\beta\,dx^i$
as totally independent gravitational `potentials':
$(g_{\alpha\beta},\vartheta^\alpha,\Gamma_\alpha{}^\beta)$. For the
metric-affine gauge theory of gravity (MAG), which encompasses the
Poincar\'e gauge theory, the Einstein-Cartan theory, and general
relativity as specific subcases, we developed a couple of computer
algebra programs which were and are used in our respective groups for
the search of exact solutions, e.g. We would like to explain and
demonstrate them. They were developed over the last eight years or so
and different people were involved: Dermott McCrea$^\dagger$, Werner
Esser, Frank Gronwald, Ralf Hecht, Yuri Obukhov, Roland Puntigam,
Sergei Tertychniy, Romualdo Tresguerres, Eugen Vlachynsky, and, more
recently, Jos\'e Socorro.

We will proceed as follows: The general metric-affine framework,
including all the conventions, will be taken from \cite{PRs}, see also
\cite{Erice95}. Our computer algebra programs use Cartan's {\em
  calculus of exterior differential forms} as implemented on the
computer by means of Hearn's computer algebra system Reduce
\cite{Hearn} and Schr\"ufer's corresponding Excalc package
\cite{Schruefer}.  We will always have in mind the special case of
general relativity, i.e., we can specialize to general relativity if
desirable. This is important if one wants to extract the {\em
  post-}Riemannian structures from the computed expressions, i.e., the
{\em deviations} from the corresponding Riemannian spacetime. We are
trying to program our MAG procedures as some kind of master programs
which can be immediately and efficiently specialized to Riemann-Cartan
or Riemann spacetimes. 

In this sense, we turn, in Sec.\ref{curv}, to the {\em curvature} 2-form of
a metric-affine space with its $16\times 16$ independent components.
We will {\em irreducibly decompose} it under the Lorentz group and discover
thereby typical post-Riemannian pieces, besides the pieces which
emerge already in a Riemannian space. This automatized irreducible
decomposition of the curvature may also be useful for people who are
not concerned with gravitational gauge theories but just with
non-Riemannian structures in {\em differential geometry}, i.e., we
believe that Sec.\ref{curv} should be of more general interest than
the rest of the paper.

In Sec.3, as an example for testing our methods, we turn to an exact
(electrovac) solution of the Einstein-Maxwell equations of Petrov type
D as formulated by {\em Pleba\'nski and Demia\'nski (P\&D)} \cite{pd},
see also Garc\'{\i}a D\'{\i}az \cite{Alberto} and Debever, Kamran, and
McLenaghan \cite{Debever0,Debever}. After a short review of the
Einstein-Maxwell equations in exterior calculus, we develop the
computer code for the orthonormal coframe of the P\&D solution and
compute its {\em curvature}.

Subsequently, we will determine the electromagnetic energy-momentum
distribution of the P\&D solution. For this purpose we bring in our
{\em Maxwell program code} which is suitable for electromagnetic
fields on top of an arbitrary differentiable manifold, whatever metric
and whatever connection, see \cite{PuntigamSH,PLH97}. This so-called
metric-free formulation of Maxwell's equations (the metric enters only
via the Hodge star into the constitutive assumption) is extremely
concise and well adapted to executing such computations within
geometrical theories of gravitation.

We combine the decomposition method of Sec.2 with this Maxwell
program. We can now easily verify the Einstein-Maxwell equations and
find, in addition, a very compact form for Weyl's conformal curvature
2-form.

In Sec.4, we turn our attention again to non-Riemannian geometries. We
describe programs for the {\em irreducible decomposition} of {\em
  torsion} and {\em nonmetricity}.

In Sec.5, the P\&D metric of GR is extended to an {\em exact solution}
in the framework of {\em metric-affine gravity}. For that purpose a
non-Riemannian covector triplet has to be introduced, see \cite
{Tres1,Tres2,TuckerWang,OVETH,VTOH,TuckO,PLH97,OVEH,TuckerJadwisin,jadwisin}.
Using our decomposition programs as described in Secs.2 and 4, we are
able to test this whole geometro-physical construct provided we also
use our Maxwell package.  It should be pointed out that this exact
solution of metric-affine gravity was only found by means of an
extensive use of computer algebra.  In Sec.6 we will conclude with a
discussion.

\section{Curvature in non-Riemannian and Riemannian spaces, its irreducible 
  decomposition by machine}\label{curv}

Before we turn to the programs, let us report on the irreducible
decomposition of the curvature 2-form
\begin{equation}
  R_\alpha{}^\beta = d \Gamma_\alpha{}^\beta -\Gamma_\alpha{}^\gamma
  \wedge \Gamma_\gamma{}^\beta = \frac{1}{2}\,R_{\mu \nu \alpha }{}
  ^\beta\, \vartheta^\mu \wedge \vartheta^\nu= \frac{1}{2}\,R_{ij
    \alpha }{} ^\beta\, dx^i\wedge dx^j \,,\label{curvature}
\end{equation}
see \cite{PRs} for a discussion and the relevant literature. Besides
the linear connection $\Gamma_\alpha{}^\beta $, we have a metric $g$
available with signature $(-+++)$. We may lower the second index on
the curvature two-form. We consider the irreducible decomposition of
$R_{\alpha \beta}$ in 4 dimensions under the Lorentz group. The first
step is to separate it into its antisymmetric and symmetric parts,
\begin{equation}\label{curvdec}
  R_{\alpha \beta} =W_{\alpha \beta} + Z_{\alpha
    \beta}\;\sim\; {\tt curv2(-a,-b):=w2(-a,-b)+z2(-a,-b)}\,,
\end{equation}
with
\begin{equation}\label{vortex}
  W_{\alpha \beta} :=R_{[\alpha \beta]}\;\sim\; {\tt w2(-a,-b):=
    (curv2(-a,-b)-curv2(-b,-a))/2}
\end{equation}
and
\begin{equation}\label{strain}
  Z_{\alpha \beta} :=R_{[\alpha \beta]}\;\sim\;{\tt
    z2(-a,-b):=(curv2(-a,-b)+curv2(-b,-a))/2}\,.
\end{equation}
We displayed at the same time the computer names (identifiers) of the
quantities emerging and the formulae in Reduce-Excalc syntax. A
tutorial on the use of Excalc is contained in \cite{Stauffer}, e.g.
We recall that lower indices are denoted by a minus sign. For
convenience, in Excalc, we use Latin indices for the frames instead of
Greek ones as in the formulae. The rank $p$ of a form (if $p>0$) is
usually indicated by a corresponding Arabic number in the computer
name of the quantity, that is, the curvature is a 2-form, thus we call
it {\tt curv2}. The Einstein form is a 3-form or {\tt einstein3(a)},
etc.

\subsection{Rotational curvature}

The antisymmetric piece $W_{\alpha \beta}$, the `rotational
curvature', can be decomposed, similarly as in a Riemann-Cartan
spacetime, into 6 pieces,
\begin{eqnarray}
W_{\alpha \beta} &=& {^{(1)}W}_{\alpha \beta} +  {^{(2)}W}_{\alpha \beta} 
+  {^{(3)}W}_{\alpha \beta} +  {^{(4)}W}_{\alpha \beta} 
+  {^{(5)}W}_{\alpha \beta} +  {^{(6)}W}_{\alpha \beta} \nonumber\\
&\sim& {\tt weyl + paircom + pscalar + ricsymf + ricanti + scalar}\,.
\end{eqnarray}
Here {\tt weyl} corresponds to $ {^{(1)}W}_{\alpha \beta}$ etc.

We can `contract' the 2-form $W_{\alpha\beta}$ in different ways by
means of the frame $e_\alpha\sim{\tt e(-a)}$, the coframe
$\vartheta^\alpha\sim{\tt o(a)}$, and the Hodge star $^\star\sim\#$.
The corresponding truncated quantities read:
\begin{eqnarray}
  W^\alpha\,& :=&\, e_\beta \rfloor W^{\alpha
    \beta}\;\sim\;\qquad\quad{\tt wone1(a):=e(-b)\_\, | w2(a,b)}\,,\\ 
  W\,&:=&\, e_\alpha \rfloor W^\alpha\;\sim\;\;\;\,\qquad\qquad{\tt
    wzero:=e(-a)\_\, | wone1(a)}\,,\\ X^\alpha\,&:=&\, {^\star
    (}W^{\beta\alpha} \wedge \vartheta_\beta)\;\sim\;\quad{\tt xone1(a):=}
   \; \#{\tt (w2(b,a)\wedge o(-b))}\,,\\ X\,&:=&\, e_\alpha \rfloor
  X^\alpha\;\sim\,\qquad\qquad\quad{\tt xzero:=e(-a)\_\, | xone1(a)}\,.
\end{eqnarray}
They can further be decomposed into
\begin{eqnarray}
\Psi_\alpha &:=& X_\alpha - \frac{1}{4} \vartheta_\alpha \wedge X -
\frac{1}{2} e_\alpha \rfloor (\vartheta^\beta \wedge X_\beta)\,, \\
\Phi_\alpha &:=& W_\alpha - \frac{1}{4} W \vartheta_\alpha - \frac{1}{2}
e_\alpha \rfloor(\vartheta^\beta \wedge W_\beta)\,,
\end{eqnarray}
or
\begin{verbatim}
psi1(-a) := xone1(-a)-(1/4)*o(-a)^xzero
                   -(1/2)*e(-a)_|(o(b)^xone1(-b))$
phi1(-a) := wone1(-a)-(1/4)*wzero*o(-a)
                   -(1/2)*e(-a)_|(o(b)^wone1(-b))$
\end{verbatim}
and then `blown up' to the 2-forms
\begin{eqnarray}\label{paircom}
  {\tt paircom2(-a,-b)}\;&\sim &\;- {^\star (}\vartheta_{[\alpha}\wedge
  \Psi_{\beta]})\,, \\ {\tt pscalar2(-a,-b)}\;&\sim &\;- \frac{1}{12}
  {^\star (}X \wedge \vartheta_\alpha \wedge \vartheta_\beta)\,, \\ {\tt
    ricsymf2(-a,-b)}\;&\sim &\; - \vartheta_{[\alpha}\wedge
  \Phi_{\beta]}\,, \\ {\tt ricanti2(-a,-b)}\;&\sim &\;-\frac{1}{2}
  \vartheta_{[\alpha}\wedge e_{\beta]}\rfloor (\vartheta^\gamma \wedge
  W_\gamma)\,, \\ {\tt scalar2(-a,-b)}\; &\sim &\;-\frac{1}{12} W
  \vartheta_\alpha \wedge \vartheta_\beta\,,\\ {\tt weyl2(-a,-b)}\;&\sim &\;
  W_{\alpha \beta} - {\sum_{I=2}^6} {^{(I)}W}_{\alpha \beta}\, .
\end{eqnarray}
To be concrete, let us display (\ref{paircom}), e.g., as a
Reduce-Excalc command:
\begin{verbatim}
     paircom2(-a,-b) :=-(1/2)*#(o(-a)^psi1(-b)-o(-b)^psi1(-a)); 
\end{verbatim}
The hash sign {\tt \#} represents the Hodge star operator.

Analogously, we can also translate the remaining definitions of the
irreducible components into Reduce-Excalc. We remember that the rank
of a form in Reduce-Excalc has to be declared explicitly by means of
the {\tt pform} declaration. In this way we arrive at the
decomposition file:

\begin{verbatim}
%***************************************************************
%  Irred. decomposition of the rotational curvature w2(-a,-b)  *
%***************************************************************
% file rotcurvdecomp.exi, 1998-03-15
% input: coframe o(a), frame e(a), metric g(a,b), curv2(a,b)

pform {wzero,xzero}=0,
      {wone1(a),xone1(a),psi1(a),phi1(a)}=1,
      {w2(a,b),paircom2(a,b),pscalar2(a,b),ricsymf2(a,b),
               ricanti2(a,b), scalar2(a,b),   weyl2(a,b)}=2$

w2(-a,-b) := (1/2)*(curv2(-a,-b)-curv2(-b,-a))$
wone1(a)  := e(-b)_|w2(a,b)$
wzero     := e(-b)_|wone1(b)$
xone1(a)  := #(w2(b,a)^o(-b))$
xzero     := e(-a)_|xone1(a)$
psi1(-a)  := xone1(-a)-(1/4)*o(-a)^xzero
                      -(1/2)*e(-a)_|(o(b)^xone1(-b))$
phi1(-a)  := wone1(-a)-(1/4)*o(-a)*wzero
                      -(1/2)*e(-a)_|(o(b)^wone1(-b))$

paircom2(-a,-b) := -(1/2)*#(o(-a)^psi1(-b)-o(-b)^psi1(-a));
pscalar2(-a,-b) := -(1/12)*#(xzero^o(-a)^o(-b));
ricsymf2(-a,-b) := -(1/2)*(o(-a)^phi1(-b)-o(-b)^phi1(-a));
ricanti2(-a,-b) := -(1/4)*(o(-a)^(e(-b)_|(o(c)^wone1(-c)))
                          -o(-b)^(e(-a)_|(o(c)^wone1(-c)))); 
scalar2(-a,-b)  := -(1/12)*wzero*o(-a)^o(-b);
weyl2(-a,-b)    := w2(-a,-b)-paircom2(-a,-b)-pscalar2(-a,-b)-
                 ricsymf2(-a,-b)-ricanti2(-a,-b)-scalar2(-a,-b);

clear wzero,xzero,wone1(a),xone1(a),psi1(a),phi1(a)$
$end$
%***************************************************************
\end{verbatim}

Before this program segment can be used, we have to specify a coframe
{\tt o(a)}, the frame {\tt e(-a)} dual to it, a metric {\tt g}, and a
connection {\tt gamma(-a,b)}, and to compute the curvature {\tt
  curv2(-a,b)} beforehand. This will be done below.

Already in a Riemann-Cartan space (with metric compatible connection),
i.e., in the case of vanishing nonmetricity, there emerge these 6
pieces of the curvature tensor with their $10+9+1+9+6+1$ independent
components. In a {\em Riemannian space}, we have {\tt paircom =
  pscalar = ricanti = 0}, or
\begin{equation}
  R_{\alpha\beta\gamma\delta}=R_{\gamma\delta\alpha\beta}\,,\quad
  R_{[\alpha\beta\gamma\delta]}=0\,,\quad
  Ric_{[\alpha\beta]}=0\,,
\label{riem2}\end{equation}
with $Ric_{\alpha\beta}:=R_{\gamma\alpha\beta}{}^\gamma$.  Then only
{\tt weyl}, {\tt ricsymf} ({\tt ric}ci {\tt sym}metric and trace{\tt
  f}ree), and {\tt scalar} survive with their, as is well known
from general relativity, $10+9+1$ independent components. In other
words, the torsion $T^\alpha$, via the first Bianchi identity
$DT^\alpha=R_\beta{}^\alpha\wedge\vartheta^\beta$, induces the
{\em non-}vanishing of {\tt paircom}, {\tt pscalar}, and {\tt ricanti}.
Obviously, (\ref{riem2}) represents the $16$ algebraic identities of a
Riemannian curvature tensor.

\subsection{Strain curvature = segmental curvature 
  $\oplus$ shear curvature}

The symmetric part $Z_{\alpha \beta}$ of the curvature two-form, the
`strain curvature', is more involved than the rotational curvature.
First of all, we split it into a tracefree (or shear) and a trace (or
dilation) part:
\begin{equation}
  Z_{\alpha \beta} = {\nearrow\!\!\!\!\!\!\! Z}_{\alpha \beta} +
  \frac{1}{4}\,g_{\alpha \beta}\, Z\,, \qquad Z:= Z_\gamma{}^\gamma\,.
\end{equation}
The dilation piece $g_{\alpha \beta} Z_\gamma{}^\gamma/4\,$ is Weyl's
{\em segmental curvature}, which we recognize as one irreducible piece of
the strain curvature. The rest, the shear curvature
${\nearrow\!\!\!\!\!\!\! Z}_{\alpha \beta}$, will be `contracted':
\begin{equation}{\nearrow\!\!\!\!\!\!\! Z}_\alpha:=e^\beta\rfloor
  {\nearrow\!\!\!\!\!\!\! Z}_{\alpha\beta}\,, \qquad
  \Delta:=\frac{1}{2}\,(\vartheta^\alpha\wedge {\nearrow\!\!\!\!\!\!\!
    Z}_\alpha)\,,\qquad Y_\alpha:=\,^\star ({\nearrow\!\!\!\!\!\!\!
    Z}_{\alpha\beta}\wedge\vartheta^\beta )\,.\end{equation} An even
finer decomposition is possible:
\begin{equation}\Xi_\alpha:=  {\nearrow\!\!\!\!\!\!\! Z}_\alpha 
  -\frac{1}{2}e_\alpha\rfloor (\vartheta^\gamma
  \wedge{\nearrow\!\!\!\!\!\!\! Z}_\gamma)\,, \qquad\qquad
  \Upsilon_\alpha:= Y_\alpha-
  \frac{1}{2}\,e_\alpha\rfloor(\vartheta^\gamma\wedge
  Y_\gamma)\,.\end{equation}

In terms of these newly introduced quantities, the irreducible
decomposition of $Z_{\alpha \beta}$ may be written as
\begin{eqnarray}
  Z_{\alpha \beta}&=&{^{(1)}Z}_{\alpha \beta}+{^{(2)}Z}_{\alpha \beta}
  +{^{(3)}Z}_{\alpha \beta}+{^{(4)}Z}_{\alpha \beta}+{^{(5)}Z}_{\alpha
    \beta}, \\ &\sim&{\tt zcurvone +zcurvtwo+zcurvthree
    +dilcurv+zcurvfive\,,}\nonumber 
\end{eqnarray}
where
\begin{eqnarray}
  {\tt zcurvtwo2(-a,-b)}\;&\sim&\; -\frac{1}{2}\; ^\star\left\{
    \vartheta_{(\alpha}\wedge \Upsilon_{\beta)} \right\}, \\ {\tt
    zcurvthree2(-a,-b)}\;&\sim&\; \frac{1}{3}\, \left\{ 2
    \vartheta_{(\alpha} \wedge (e_{\beta)} \rfloor \Delta ) -
    g_{\alpha \beta} \Delta \right\}, \\ {\tt dilcurv2(-a,-b)}
  \;&\sim&\; \frac{1}{4}\;g_{\alpha \beta} \,Z\,, \\ {\tt
    zcurvfive2(-a,-b)}\;&\sim&\; \frac{1}{2}\; \vartheta_{(\alpha}
  \wedge \Xi_{\beta)}\,, \\ {\tt zcurvone2(-a,-b) }\;&\sim&\; Z_{\alpha
    \beta} - {\sum_{I=2}^5} \, {^{(I)}Z}_{\alpha \beta}\,.
\end{eqnarray}
The corresponding Reduce-Excalc code reads as follows:

\begin{verbatim}
%***************************************************************
%   Irred. decomposition of the strain curvature z2(-a,-b)     *
%***************************************************************
% file straincurvdecomp.exi, 1998-03-15
% input: coframe o(a), frame e(a), metric g(a,b), curv2(a,b)

pform {ztracef1(a),yy1(a),xi1(a),upsilon1(a)}=1,
      {z2(a,b),ztracef2(a,b),delta2,   
       zcurvone2(a,b), zcurvtwo2(a,b),zcurvthree2(a,b),
        dilcurv2(a,b),zcurvfive2(a,b)}=2$

z2(-a,-b)       := (1/2)*(curv2(-a,-b)+curv2(-b,-a))$
ztracef2(-a,-b) := z2(-a,-b)-(1/4)*g(-a,-b)*z2(-c,c)$
ztracef1(-a)    := e(b)_|ztracef2(-a,-b)$
delta2          := (1/2)*o(a)^ztracef1(-a)$
yy1(-a)         := #(ztracef2(-a,-b)^o(b))$
xi1(-a)        := ztracef1(-a)-(1/2)*e(-a)_|(o(c)^ztracef1(-c))$ 
upsilon1(-a)    := yy1(-a)-(1/2)*e(-a)_|(o(c)^yy1(-c))$

zcurvtwo2(-a,-b)  := -(1/4)*#(o(-a)^upsilon1(-b)
                             +o(-b)^upsilon1(-a));
zcurvthree2(-a,-b):= (1/6)*(2*(o(-a)^(e(-b)_|delta2)
                     +o(-b)^(e(-a)_|delta2))-2*g(-a,-b)*delta2);
dilcurv2(-a,-b)   := (1/4)*g(-a,-b)*z2(-c,c);
zcurvfive2(-a,-b) := (1/4)*(o(-a)^xi1(-b)+o(-b)^xi1(-a));
zcurvone2(-a,-b)  := z2(-a,-b)-zcurvtwo2(-a,-b)
          -zcurvthree2(-a,-b)-dilcurv2(-a,-b)-zcurvfive2(-a,-b);

clear ztracef1(a),yy1(a),xi1(a),upsilon1(a),
      ztracef2(a,b),delta2$   
$end$
%***************************************************************
\end{verbatim}

\section{Intermezzo: Einstein-Maxwell equations, Pleba\'nski 
\& Demia\'nski metric}

Let us now turn to general relativity (GR) in order to have a decent
electrovac solution of GR as a starting point for our further
considerations, i.e.\ an exact solution of the Einstein equation with
a distribution of electromagnetic energy-momentum as source. The
Einstein-Maxwell equations with cosmological constant $\lambda$ read:
\begin{equation}\label{Einstein}
  G_\alpha+\lambda\,\eta_\alpha=\;2\stackrel{{\rm Max}}{T_\alpha}\,, \qquad
  dH=0\,,\qquad dF=0\,,\qquad H=\,^\star F\,.\end{equation} The
Einstein 3-form $G_\alpha\sim{\tt einstein3(-a)}$ and the
electromagnetic energy-momentum 3-form
$\stackrel{{\rm Max}}{T_\alpha}\sim{\tt maxenergy3(-a)}$ are defined by
\begin{equation}\label{Eindef}
G_\alpha:=\frac{1}{2}\,\eta_{\alpha\beta\gamma}\wedge R^{\beta\gamma}
\end{equation} and
\begin{equation}\label{maxenergy}
  \stackrel{{\rm Max}}{T_\alpha}=e_\alpha\rfloor L_{{\rm
      Max}}+(e_\alpha\rfloor F)\wedge H\,,\qquad {\rm with}\qquad
  L_{{\rm Max}}:=-\frac{1}{2}\,F\wedge H \,,\end{equation}
respectively. The excitation $H\sim{\tt excit2}\,$ and the
electromagnetic field strength $F\sim{\tt farad2}\,$ are given by
\begin{equation}
H= - {\cal H}\, \wedge \, dt + {\cal D}\,,\qquad F=  E\wedge dt + B\,
\end{equation}($ {\cal H}$ magnetic and ${\cal D}$ electric 
excitation, $E$ electric and $B$ magnetic field).  The ubiquitous
$\eta$-basis is dual to the $\vartheta$-basis, that is, see
\cite{PRs},
\begin{equation}\label{eta}
  \eta_\alpha=\,^\star\vartheta_\alpha\,,\qquad
  \eta_{\alpha\beta}=\,^\star\left(\vartheta_\alpha\wedge
    \vartheta_\beta\right)\,,\qquad{\rm etc.}\end{equation}

In a {\em Riemannian} space, we can decompose the Einstein 3-form into
2 irreducible pieces\footnote{In a metric-affine space, the definition
  (\ref{Eindef}) kills the strain curvature and also 3 of the
  rotational curvature pieces. We find:\begin{equation}
    G_\alpha=\frac{1}{2}\,\eta_{\alpha\beta\gamma}\wedge\sum_{I=4}^{6}
    {^{(I)}}W^{\beta\gamma}\,.\end{equation} Consequently, in a
  metric-affine space, besides {\tt ricsymf} and {\tt scalar}, only
  {\tt ricanti} surfaces as an additional quantity in the Einstein
  3-form.}:
\begin{eqnarray}
  G_\alpha&=&\frac{1}{2}\,\eta_{\alpha\beta\gamma}\wedge
  \left[^{(4)}R^{\beta\gamma}+\,^{(6)}R^{\beta\gamma}\right]\nonumber
  \\ &=&\frac{1}{2}\,\eta_{\alpha\beta\gamma}\wedge
  \,^{(4)}R^{\beta\gamma} -\frac{1}{4}\,R\,\eta_\alpha \,,\qquad{\rm
    with}\qquad R:=e_\alpha\rfloor e_\beta\rfloor R^{\alpha\beta}\,.
\end{eqnarray}If we substitute this into the Einstein equation in 
(\ref{Einstein}), then it splits into two independent pieces, a
tracefree one and the trace:
\begin{equation}\label{Einsplit}
  \frac{1}{2}\,\eta_{\alpha\beta\gamma}\wedge
  \,^{(4)}R^{\beta\gamma}\,=\; 2\stackrel{{\rm Max}}{T_\alpha}\,,\qquad
  R=4\lambda\,.
\end{equation}
Note that $\stackrel{{\rm Max}}{T_\alpha}$ is tracefree:
$\vartheta^\alpha \wedge \stackrel{{\rm Max}}{T_\alpha}=0$. It is now
evident of how we can directly use our decomposition program of the
last section in order to verify the field equations (\ref{Einsplit}).
But, of course, we could also compute the Einstein 3-form directly
according to its definition (\ref{Eindef}).

\subsection{P\&D coframe {\tt o(a)} and electromagnetic 
  potential {\tt pot1}}

The P\&D solution is described in terms of coordinates
$x^\mu=(\tau,q,p,\sigma)\sim{\tt (tau,q,p,sigma)}$. We use Eq.(3.30)
of Plebanski and Demianski \cite{pd}, see also
\cite{Alberto,Debever0,Debever,blackhole}.  Then the {\em orthonormal}
coframe reads,
\begin{eqnarray}
  \vartheta^{\hat{0}} &=&\frac{1}{H}\sqrt{\frac{{\cal Q}}{\Delta} }
  \left(d\tau - p^2 d\sigma \right)\,, \nonumber\\ \vartheta^{\hat{1}}
  &=&\frac{1}{H}\sqrt{\frac{\Delta}{{\cal Q}} }\; dq,\nonumber\\ 
  \vartheta^{\hat{2}} &=&\frac{1}{H}\sqrt{\frac{\Delta}{{\cal P}} }\;
  dp, \nonumber\\ \vartheta^{\hat{3}} &=&\frac{1}{H}\sqrt{\frac{{\cal
        P}}{\Delta} } \left(d\tau + q^2 d\sigma \right)\,, \label{U}
\end{eqnarray}
with the metric
\begin{equation} g=-\vartheta^{\hat{0}}\otimes \vartheta^{\hat{0}}
  +\vartheta^{\hat{1}}\otimes \vartheta^{\hat{1}}
  +\vartheta^{\hat{2}}\otimes \vartheta^{\hat{2}}
  +\vartheta^{\hat{3}}\otimes
  \vartheta^{\hat{3}}\,,\label{localmetric}\end{equation}
or
\begin{equation}
  g= \frac{1}{H^2}\left\{- \frac{{\cal Q}}{\Delta}\; (d \tau - p^2
    d\sigma )^2 + \frac{\Delta}{{\cal Q}}\; dq^2 +\frac{\Delta}{{\cal
        P}}\; dp^2 +\frac{{\cal P}}{\Delta}\; (d \tau + q^2 d
    \sigma)^2 \right\}\,.
\label{pd}
\end{equation}
The functions ${\cal P}, {\cal Q},\Delta$, and $H$ are polynomial in
$p$ and $q$. With the parameters $m,n,e_{\rm o},g_{\rm
  o},b,\epsilon,\mu$, they read:
\begin{eqnarray}
  {\cal P} &:=& \left(b - g_{\rm o}^2 \right) + 2np - \epsilon p^2 + 2
  m \mu p^3 - \left[\mu^2 \left(b + e_{\rm o}^2 \right) +
    \frac{\lambda}{3}\right] p^4 ,\nonumber\\ {\cal Q} &:=& \left(b +
    e_{\rm o}^2\right) - 2mq + \epsilon q^2 - 2n\mu q^3 - \left[ \mu^2
    \left(b -g_{\rm o}^2\right) + \frac{\lambda}{3} \right] q^4 ,
  \nonumber \\ \Delta &:=& p^2 + q^2 ,\nonumber \\ H &:=& 1 - \mu \, p
  \, q\,.
\label{solu}
\end{eqnarray}
We recover the original version of P\&D \cite{pd}, if we put
$b:=\gamma - \lambda/6$ and $\mu=1$. The parameter $\mu$, see
\cite{blackhole}, generalizes the P\&D solution slightly.

The electromagnetic potential $A$ appropriate for this solution can be
expressed as follows, see \cite{pd} ($e_{\rm o}$ = electric and
$g_{\rm o}$ = magnetic charge):
\begin{eqnarray}
  { A}&=& \frac{1}{\Delta}\left[( e_{\rm o}\,q + g_{\rm o}\,p)\, { d
      \tau} + (g_{\rm o}\,q - e_{\rm o}\,p)\, p \,q \,{ d \sigma} \right
  ]\nonumber\\ &=& \frac{H}{\sqrt{\Delta}}\left( \frac{e_{\rm o}\, q
      }{\sqrt{{\cal Q}}} \;\vartheta^{\hat{0}} + \frac{g_{\rm o}\,
      p}{\sqrt{{\cal P}}} \;\vartheta^{\hat{3}}\right)\,.
\label{W}
\end{eqnarray}

In the following we will show of how one can put this metric into
Excalc.  In an interactive Reduce 3.6 session, the Excalc package can
be loaded by the command {\tt load\_package excalc\$} . In the next
step, Excalc has to be acquainted with the rank of the differential
forms to be used in the code. This is achieved by the {\tt pform}
declaration.  Subsequently it has to be specified by means of {\tt
  fdomain} on which variables these forms do depend. Afterwards one
defines the coframe:
\begin{verbatim}
%***************************************************************
%                 Specifying the coframe o(a)                  *
%***************************************************************
% file coframe.exi, 1998-03-15
% no prior input

load_package excalc$

pform   {hh,sqrtqq,sqrtpp,delta,polynomq,polynomp}=0$ 
fdomain hh=hh(p,q),sqrtqq=sqrtqq(q),sqrtpp=sqrtpp(p),delta=
        delta(p,q),polynomq=polynomq(q),polynomp=polynomp(p)$

coframe o(0) = ((1/hh)*sqrtqq/sqrt(delta))*(d tau-p**2*d sigma), 
        o(1) = ((1/hh)*sqrt(delta)/sqrtqq)* d q,             
        o(2) = ((1/hh)*sqrt(delta)/sqrtpp)* d p,
        o(3) = ((1/hh)*sqrtpp/sqrt(delta))*(d tau+q**2*d sigma)  
with 
    metric g =  -o(0)*o(0) + o(1)*o(1) + o(2)*o(2) + o(3)*o(3)$
frame e$
$end$
%***************************************************************
\end{verbatim}
Here the frame $e_\alpha\sim{\tt e(-a)} $ is dual to the coframe.

In accordance with (\ref{solu}), we introduced the functions:
\begin{eqnarray}
{\tt hh}\;&\sim&\; {1- \mu\, p\,q} \,,\nonumber\\
{\tt delta}\;&\sim&\; {{p^2+ q^2}}\,, \nonumber\\
{\tt sqrtqq}\;&\sim&\; \sqrt {\cal Q}\,, \nonumber\\
{\tt sqrtpp}\;&\sim&\; \sqrt {\cal P}\,.
\end{eqnarray}
Here $\cal Q$ and $\cal P$, see (\ref{solu}), are polynomials of
quartic order in {\tt q} and {\tt p}, respectively. For the time
being, we will only read in $\Delta$ and $H$, since they are not very
complicated:
\begin{verbatim}
delta := p**2 + q**2;
hh    :=    1 - mu*p*q;
\end{verbatim}
However, the explicit form of $\sqrt{\cal Q}$ and $\sqrt{\cal P}$ will
be left open.

It is an old trick to treat square roots of functions in the way we
did it. Bringing in square roots explicitly, will usually slow down a
program appreciably or even may blow it up beyond the capacity of the
computer. We can express the derivatives of $\sqrt{\cal Q}$ and
$\sqrt{\cal P}$ by means of the chain rule, also their squares get a
new name:
\begin{verbatim}
@(sqrtpp,p) := @(polynomp,p)/(2*sqrtpp); 
@(sqrtqq,q) := @(polynomq,q)/(2*sqrtqq);

sqrtpp**2   := polynomp;
sqrtqq**2   := polynomq;
\end{verbatim}

\subsection{The curvature {\tt riem2(a,b)} of the P\&D solution}

Let us first bring in the metric volume element and the $\eta$-basis:

\begin{verbatim}
%***************************************************************
%     The metric volume element eta4 and the eta basis         *
%***************************************************************
% file eta.exi, 1998-03-15
% prior input: coframe o(a), frame e(a), metric g(a,b)

pform eta0(a,b,c,d)=0,eta1(a,b,c)=1,eta2(a,b)=2,eta3(a)=3, eta4=4$

eta4          := # 1$
eta3(a)       := e(a) _| eta4$
eta2(a,b)     := e(b) _| eta3(a)$
eta1(a,b,c)   := e(c) _| eta2(a,b)$
eta0(a,b,c,d) := e(d) _| eta1(a,b,c)$
$end$
%***************************************************************
\end{verbatim}
This is simple geometry. Now we compute connection and curvature:

\begin{verbatim}
%***************************************************************
%   Riemannian connection chris1(a,b) and curvature riem2(a,b) *
%***************************************************************
% file riemann.exi, 1998-03-15
% prior input; o(a), e(a), g(a,b) 

factor ^,o(0),o(1),o(2),o(3)$ 
 
Riemannconx chris1$
% because of our conventions, we have to take the transp. of it

chris1(-a,b):=chris1(b,-a)$
 
pform {riem2(a,b),curv2(a,b)}=2, {einstein3(a),cosmolog3(a)}=3,
      grscalar=0$
% antisymmetric riem2; since we want to turn non-Riem. later
 
riem2(-a,b)  := d chris1(-a,b) - chris1(-a,c) ^ chris1(-c,b)$
grscalar     := e(-a) _| (e(-b) _| riem2(a,b))$

einstein3(a) := (1/2) * eta1(a,b,c) ^ riem2(-b,-c);
cosmolog3(a) := lam * eta3(a)$

curv2(a,b)   := riem2(a,b)$
$end$
%***************************************************************
\end{verbatim}

\subsection{Starting an interactive Reduce session}

Let us start such a session by calling Reduce and reading in the
coframe and the $\eta$ files. Then we specify $\Delta$, $H$, etc.\ and
call the Riemann file:
\begin{verbatim}
reduce
in "coframe.exi"$
in "eta.exi"$
delta := p**2 + q**2;
hh    :=    1 - mu*p*q;

@(sqrtpp,p) := @(polynomp,p)/(2*sqrtpp); 
@(sqrtqq,q) := @(polynomq,q)/(2*sqrtqq);

sqrtpp**2   := polynomp;
sqrtqq**2   := polynomq;

in "riemann.exi"$
\end{verbatim}
We hasten to point out that we copied the commands listed above from
the original tex-file of this paper directly into a separate Reduce
window.  Thereby our Reduce-Excalc code is on line computer checked
and, thus, absolutely trustworthy.

After a few seconds (with a Sun sparcstation) the curvature will be in
the output in terms of the up to now unspecified functions {\tt
  sqrtqq, sqrtpp, polynomq, polynomp}.  Nevertheless, if we read in
the rotational and the strain curvature decomposition programs of
above,
\begin{verbatim}
in "rotcurvdecomp.exi"$
in "straincurvdecomp.exi"$
\end{verbatim}
we immediately find {\tt weyl2, ricsymf2, scalar2} as non-vanishing,
whereas the whole rest vanishes, as it must be in a Riemannian
spacetime. Before we look into more details of the surviving curvature
pieces, we discuss the coupling to the electromagnetic field.

\subsection{The electromagnetic field as source of gravity: 
  {\tt maxenergy3(a)}}

Our Maxwell program has been published earlier, see \cite{PuntigamSH}.
It has to be preceded by the coframe statement and the specification
of the frame {\tt e(a)}. This has been already executed in the last
subsection. Therefore we can take the electromagnetic potential from
(\ref{W}) and type it in directly into the Maxwell program:

\begin{verbatim}
%***************************************************************
%             Maxwell equations metric free                    *
%***************************************************************
% file maxwell.exi, 1998-03-15
% prior input: coframe o(a), frame e(a), metric g(a,b) 

% potential is prescribed
% field strength farad2, excitation excit2, and the left hand 
% sides of the Maxwell equations are defined

pform pot1=1, {farad2,excit2}=2, {maxhom3,maxinh3}=3$

pot1    := (1/delta)*(  (ee*q+gg*p)     * d tau 
                      + (gg*q-ee*p)*p*q * d sigma); 
farad2  := d pot1;
maxhom3 := d farad2;
excit2  := # farad2;
maxinh3 := d excit2;
 
% Maxwell Lagrangian and energy-momentum current are assigned

pform lmax4=4, maxenergy3(a)=3$

lmax4          := -(1/2) * farad2 ^ excit2;
maxenergy3(-a) := e(-a) _| lmax4 + (e(-a) _| farad2) ^ excit2;
$end$
%***************************************************************
\end{verbatim}

In continuing our interactive Reduce session, we read in this file:
\begin{verbatim}
in "maxwell.exi"$
\end{verbatim}
Both Maxwell equations are fulfilled. The electromagnetic
energy-momentum is determined. If we type in
\begin{verbatim}
pot1;
\end{verbatim}
the frame version of $A$ will be displayed, cf.\ (\ref{W}). The Einstein
equation with cosmological constant and electromagenetic field source
can be written down:
\begin{verbatim}
pform einsteinmax3(a)=3$
einsteinmax3(a):=einstein3(a) + cosmolog3(a) - 2*maxenergy3(a); 
\end{verbatim}

The left hand side, namely {\tt einsteinmax3(a)} if evaluated, has
already a very compact form.  Then, substituting the polynomials $\cal
P$ and $\cal Q$ of (\ref{solu}),
\begin{verbatim}
% substitution
polynomp := bb - gg**2  + 2*n*p  - epsi*p**2 + 2*mu*m*p**3  
           -(lam/3 + mu**2*(bb + ee**2))*p**4;

polynomq := bb + ee**2 - 2*m*q + epsi*q**2 - 2*n*mu*q**3 
           -(lam/3 + mu**2*(bb - gg**2))*q**4;

einsteinmax3(a) := einsteinmax3(a);
showtime;
\end{verbatim}
the Einstein equation turns out to be fulfilled, too. 

\subsection{Conformal curvature two-form {\tt weyl2(a,b)} 
  of the P\&D solution}

Consequently the P\&D metric is a solution of the Einstein-Maxwell
equations, indeed. The P\&D gravitational field is characterized by
the conformal curvature 2-form
\begin{equation}
  C_{\alpha\beta}:=\,
  ^{(1)}R_{\alpha\beta}=\frac{1}{2}\,C_{\mu\nu\alpha\beta}\,
  \vartheta^\mu\wedge\vartheta^\nu\,.
\end{equation}
Therefore it is desirable to put this in a compact form. Our
decomposition program ran already. Thus we only have to collect the
result by turning the expansion off and calling for the Weyl 2-form:

\begin{verbatim}
off exp$
weyl2(a,b):=weyl2(a,b);
\end{verbatim}
We find thereby (Heinicke \cite{Heinicke}):
\begin{equation}
C^{\alpha \beta} = f(\alpha, \beta) \, \left\{-A\,\vartheta ^{\alpha} \wedge
\vartheta ^{\beta} + B\,^{\star}\!\left(\vartheta ^{\alpha} \wedge \vartheta
^{\beta} \right) \right\}\,,
\end{equation} with 

\begin{eqnarray*}
\!\!\!\!\!\!\!\!\!\!\!\!A&:=& \frac{(\mu pq-1)^3}{\left( p^2+q^2\right)^3} 
         \left[ m \, \left( 3p^2- q^2\right) q
               +n \, \left(  p^2-3q^2\right) p
               -\left(e_0^2+g_0^2\right) \, (\mu pq+1)\left(p^2-q^2\right)
         \right] \\
\!\!\!\!\!\!\!\!\!\!\!\!B&:=& \frac{(\mu pq-1)^3}{\left(p^2+q^2\right)^3} 
         \left[ m \, \left( 3q^2- p^2\right) p
               -n \, \left(  q^2-3p^2\right) q
               -\left(e_0^2+g_0^2\right) \, (\mu pq+1)2pq
         \right]                
\end{eqnarray*}
and
\begin{equation}
f(\alpha, \beta)= \bigg\{ \begin{array}{rl}
                                    2 & \mbox{ for } (\alpha,\beta) \in
                                     \{(0,1),(1,0),(2,3),(3,2) \} \\
                                    -1 & \mbox{else} 
                              \end{array}
\end{equation}

Of course, the Weyl $2$--form is a real quantity. Accordingly, we can
also represent it by its self--dual part, see \cite{W+A} ($i$=
imaginary unit),
\begin{equation}
^+C_{\alpha\beta}:=\frac{1}{2}\left({C}_{\alpha\beta}
+i\,^\star{C}_{\alpha\beta} \right)=\frac{1}{2}\,^+{C}_{\mu\nu\alpha\beta}
\,\vartheta^\mu\wedge\vartheta^\nu
\label{selfdual}\,,
\end{equation} or, in computer code, by
\begin{verbatim}
on complex$
pform selfdualweyl2(a,b)=2$
selfdualweyl2(a,b) := (1/2) * (weyl2(a,b) + i * #weyl2(a,b));
\end{verbatim}
It can be very compactly displayed \cite{Heinicke},
\begin{equation}
  ^+ C^{\alpha \beta} = f(\alpha , \beta)\; {\cal C} \left\{
    \vartheta ^{\alpha} \wedge \vartheta^{\beta} +i\,^{\star}\! \left(
      \vartheta ^{\alpha} \wedge \vartheta^{\beta} \right) \right\}\,,
\end{equation}
where 
\begin{equation}
{\cal C} := -\frac{1}{2}(A+iB) = -\frac{(\mu pq-1)^3}{2(p-iq)^3}
                        \left( n-im \,-\left(e_0^2+g_0^2\right)
                                        \frac{\mu pq+1}{p+iq} \right)\,.
\end{equation}
Often it is thought that such a compact representation is only
possible in a null frame (i.e., by means of a Newman-Penrose tetrad).
Obviously, however, this is not true. It is an effect of the power of
the moving frame aspect of exterior calculus, {\em not} of that of a {\em
  null} coframe.

By the same token, we computed also the two quadratic curvature
invariants $^\star ( W_{\alpha \beta} \wedge ^\star W^{\alpha \beta})$
and $^\star ({ W_{\alpha \beta} \wedge W^{\alpha \beta}})$, but their
discussion is beyond the scope of this paper.

\section{Torsion and nonmetricity and their irreducible decompositions}

In a metric-affine spacetime, the field strengths, besides the
curvature two-form {\tt curv2(a,b)}, are given by the components of
the {\it torsion} two-form
\begin{equation}
 {\tt torsion2(a)}\sim  T^\alpha = D \vartheta^\alpha = d \vartheta^\alpha
  +\Gamma_\beta{}^\alpha \wedge \vartheta^\beta
  =\frac{1}{2}T_{\mu\nu}{}^\alpha\, \vartheta^\mu \wedge \vartheta^\nu
  \, ,
\label{torsion}
\end{equation} and the {\it nonmetricity} one-form
\begin{eqnarray} {\tt nonmet1(-a,-b)}\sim
  Q_{\alpha \beta} &=& - D g_{\alpha\beta} = - d g_{\alpha\beta} +
  \Gamma_\alpha{}^\gamma g_{\gamma\beta} + \Gamma_\beta{} \,
  ^\gamma g_{\alpha \gamma}\nonumber\\& =&\;Q_{i\alpha \beta}\, dx^i\,.
\label{nonmetricity}
\end{eqnarray}

Although torsion and nonmetricity are genuine field strengths, they can
be reinterpreted as parts of the connection. The linear connection can be
expressed in term of metric, coframe, torsion, and nonmetricity.

\subsection{Torsion {\tt torsion2(a)} and its three irreducible pieces}

For the torsion, we have the following irreducible decomposition
\begin{eqnarray}
  T^\alpha &=& {^{(1)}T}^\alpha + {^{(2)}T}^\alpha +
  {^{(3)}T}^\alpha\nonumber \\ &\sim& {\tt tentor + trator + axitor}\,,
\end{eqnarray} 
where
\begin{eqnarray}
  {\tt trator2(a)}\; &\sim&\; \frac{1}{3}\,\vartheta^\alpha \wedge
  \left(e_\beta \rfloor T^\beta\right)\,, \\ {\tt axitor2(a)}\;
  &\sim&\; \frac{1}{3}\,e^\alpha\rfloor \left(\vartheta_\beta\wedge
    T^\beta\right) \,, \\ {\tt tentor2(a)}\; &\sim&\; T^\alpha-\,
  ^{(2)}T^\alpha -\, ^{(3)}T^\alpha\,,
\end{eqnarray}or, in other `words',

\begin{verbatim}
%***************************************************************
%     Irreducible decomposition of the torsion torsion2(a)     *      
%***************************************************************
% file torsiondecomp.exi, 1998-03-15
% input: o(a), e(a), g(a,b), torsion2(a)

pform {tentor2(a),trator2(a),axitor2(a)}=2$

trator2(a):=  (1/3)*o(a)  ^ (e(-b) _| torsion2(b));
axitor2(a):=  (1/3)*e(a) _| (o(-b)  ^ torsion2(b));
tentor2(a):=   torsion2(a)-trator2(a)-axitor2(a);
$end$
%***************************************************************
\end{verbatim}

\subsection{Nonmetricity {\tt nonmet1(a,b)} and its four irreducible pieces}

The Weyl one-form $Q:= Q_\alpha{}^\alpha/4 = - g^{\alpha \beta} D
g_{\alpha \beta}/4$ is one {\it irreducible} piece of the
nonmetricity. Thus we find the traceless part of the nonmetricity as
\begin{equation}
  {\nearrow\!\!\!\!\!\!\! Q}_{\alpha \beta}:= Q_{\alpha \beta} - Q\,
  g_{\alpha \beta}\,.
\end{equation}
The traceless nonmetricity (its `deviator', in the language of
continuum mechanics) can be further `contracted':
\begin{equation}
  \Lambda_\alpha:= e^\beta \rfloor {\nearrow\!\!\!\!\!\!\! Q}_{\alpha
    \beta}\,, \quad \Lambda := \Lambda_\alpha\, \vartheta^\alpha\,,
  \nonumber
\end{equation}
\begin{equation}
  \Theta_\alpha :=\, ^\star ( {\nearrow\!\!\!\!\!\!\! Q}_{\alpha
    \beta}\wedge \vartheta^\beta )\,, \quad \Theta := \vartheta^\alpha
  \wedge \Theta_\alpha,\quad \Omega_\alpha := \Theta_\alpha -
  \frac{1}{3} \,e_\alpha \rfloor \Theta\,.
\end{equation}
From these quantities we can build up the four irreducible pieces:
\begin{eqnarray}
  {\tt binom1(-a,-b)}\;&\sim &\; \frac{2}{3} \, {^\star}
    \left(\vartheta_{(\alpha} \wedge \Omega _{\beta)}\right)\,, \\ 
  {\tt vecnom1(-a,-b)}\;&\sim &\;
  \frac{4}{9}\,\left(\vartheta_{(\beta}\,\Lambda_{\alpha)} -\frac{1}{4}\,
    g_{\alpha \beta}\, \Lambda\right)\,, \\ {\tt conom1(-a,-b)} \;&\sim
  &\; g_{\alpha \beta}\, Q\,, \\ {\tt trinom1(-a,-b)}\;&\sim &\;
  Q_{\alpha \beta}- {^{(2)}Q}_{\alpha \beta} -{^{(3)}Q}_{\alpha
    \beta}- {^{(4)}Q}_{\alpha \beta}\,.
\end{eqnarray}
{\tt trinom} represents a tensor of 3rd rank and {\tt binom} one of
2nd rank, with {\tt vecnom} and {\tt conom} we denote the covector and
the vector pieces, respectively. Clearly {\tt conom} is equivalent to
the Weyl covector.  They add up according to
\begin{eqnarray}\label{irrQ}
  Q_{\alpha \beta} &=&{^{(1)}Q}_{\alpha \beta}+ {^{(2)}Q}_{\alpha
    \beta}+ {^{(3)}Q}_{\alpha \beta}+ {^{(4)}Q}_{\alpha \beta}\\ &\sim&
  {\tt trinom + binom + vecnom + conom}\,.\nonumber
\end{eqnarray}

\begin{verbatim}
%***************************************************************
%  Irreducible decomposition of the nonmetricity nonmet1(a,b)  *
%***************************************************************
% file nonmetdecomp.exi, 1998-03-15
% input: o(a), e(a), g(a,b), nonmet1(a,b)

pform lamzero(a)=0, {weylcovector1,nomtracefree1(a,b),lamone1,
      binom1(a,b),vecnom1(a,b),trinom1(a,b),conom1(a,b)}=1,
      {thetatwo2(a),omega2(a)}=2, thetathree3=3$

weylcovector1        := nonmet1(-c,c)/4$
nomtracefree1(-a,-b) := nonmet1(-a,-b) - g(-a,-b)*weylcovector1$
lamzero(-a)          := e(b)_|nomtracefree1(-a,-b)$
lamone1              := lamzero(-a)*o(a)$
thetatwo2(-a)        := #(nomtracefree1(-a,-b)^o(b))$
thetathree3          := o(a)^thetatwo2(-a)$
omega2(-a)           := thetatwo2(-a)-(1/3)*e(-a)_|thetathree3$

binom1(-a,-b)  := (1/3)*#(o(-a)^omega2(-b)+o(-b)^omega2(-a));
vecnom1(-a,-b) := (4/9)*((o(-b)*lamzero(-a)+o(-a)*lamzero(-b))/2
                         -g(-a,-b)*lamone1/4);
conom1(-a,-b)  :=  g(-a,-b)*weylcovector1;
trinom1(-a,-b) :=  nonmet1(-a,-b) - binom1(-a,-b)
                  -vecnom1(-a,-b) - conom1(-a,-b);

clear lamzero(a),nomtracefree1(a,b),lamone1,thetatwo2(a),
      omega2(a), thetathree3=3$
$end$
%***************************************************************
\end{verbatim}

\subsection{The post-Riemannian covector triplet}

With propagating nonmetricity $Q_{\alpha\beta}$ two types of charge
are expected to arise: {\em One dilation charge} related via the
Noether procedure to the trace of the nonmetricity, the Weyl covector
$Q=Q_i \,dx^i$. It represents the connection associated with gauging
the scale transformations (instead of the $U(1)$--connection in the
case of Maxwell's field). Furthermore, {\em nine shear charges} are
expected that are related to the remaining traceless piece
${\nearrow\!\!\!\!\!\!\!Q}_{\alpha\beta}:=Q_{\alpha\beta}-
Q\,g_{\alpha\beta}$ of the nonmetricity with $4\times (4+4+1)$
components.

For the torsion and nonmetricity field configurations, we concentrate
on the simplest non--trivial case {\em with} shear. According to its
irreducible decomposition (\ref{irrQ}), the nonmetricity contains two
covector type pieces, namely $^{(4)}Q_{\alpha\beta}=
Q\,g_{\alpha\beta}$, the dilation piece, and
\begin{equation}
  ^{(3)}Q_{\alpha\beta}={4\over
    9}\left(\vartheta_{(\alpha}e_{\beta)}\rfloor \Lambda - {1\over 4}
    g_{\alpha\beta}\Lambda\right)\,,\qquad \hbox{with}\qquad \Lambda=
  \vartheta^{\alpha}e^{\beta}\rfloor\!
  {\nearrow\!\!\!\!\!\!\!Q}_{\alpha\beta}\label{3q}\,,
\end{equation}
a proper shear piece. Accordingly, our ansatz for the nonmetricity  
reads
\begin{equation}
  Q_{\alpha\beta}=\, ^{(3)}Q_{\alpha\beta} +\,
  ^{(4)}Q_{\alpha\beta}\,.\label{QQ}
\end{equation}
The torsion, in addition to its tensor piece,
encompasses a covector and an axial covector piece. Let us choose only
the covector piece as non--vanishing:
\begin{equation}
  T^{\alpha}={}^{(2)}T^{\alpha}={1\over 3}\,\vartheta^{\alpha}\wedge
  T\,, \qquad \hbox{with}\qquad T:=e_{\alpha}\rfloor
  T^{\alpha}\,.\label{TT}
\end{equation}
Thus we are left with the three non--trivial one--forms $Q$, $\Lambda$,
and $T$.  We shall assume that this {\em triplet of one--forms} shares the
spacetime symmetries, that is, its members are proportional to each other.

\section{Applying the decomposition programs in metric-affine gravity: 
  The P\&D metric together with a non-Riemannian covector triplet as
  exact solution of a dilation-shear Lagrangian}

In order to explore the potentialities of metric--affine gravity, see
\cite{Erice95}, and of the corresponding computer algebra programs, we
will choose the simple non-trivial dilation-shear gravitational
Lagrangian,
\begin{equation}
  V_{\rm dil-sh}=-\frac{1}{2l^2}\left( R^{\alpha\beta}\wedge
    \eta_{\alpha\beta}+\beta Q\wedge \hspace{-0.8em}
    {\phantom{Q}}^{\star}Q+ \gamma T\wedge \hspace{-0.8em}
    {\phantom{Q}}^{\star}T\right)-\frac{1}{8}\alpha\,
  R_{\alpha}{}^{\alpha}\wedge\hspace{-0.8em}
  {\phantom{Q}}^{\star}R_{\beta}{}^{\beta}\,,
         \label{dil-sh}
\end{equation}
with the dimensionless coupling constants $\alpha ,\;\beta$, and
$\gamma$. We will choose units such that $l^2=1$. We add the Maxwell
Lagrangian, see (\ref{maxenergy}):
\begin{equation}\label{Ltot}L=V_{\rm dil-sh}+L_{\rm Max}\,.
\end{equation} Explicitly we have:
\begin{eqnarray}\label{Ltot1}L=-\frac{1}{2l^2}\left( R^{\alpha\beta}\wedge
    \eta_{\alpha\beta}+\beta\, Q\wedge \hspace{-0.8em}
    {\phantom{Q}}^{\star}Q+ \gamma\, T\wedge \hspace{-0.8em}
    {\phantom{Q}}^{\star}T\right)&-&\frac{1}{8}\,\alpha\,
  R_{\alpha}{}^{\alpha}\wedge\hspace{-0.8em}
  {\phantom{Q}}^{\star}R_{\beta}{}^{\beta}\nonumber\\ 
  &&\quad -\frac{1}{2}\;F\wedge\hspace{-0.8em}
  {\phantom{F}}^{\star}F\,.
\end{eqnarray} 
Note the {\em formal} similarities of the curvature square Lagrangian
with the segmental curvature $\frac{1}{2}R_{\alpha}{}^{\alpha}=d\,Q$
and the Maxwell Lagrangian with the electromagnetic field strength
$F=d\,A$.

We will search for exact solutions of the field equations belonging to
this Lagrangian. Since the matter part $L_{\rm Max}$ does not depend
on the connection $\Gamma_\alpha{}^\beta$, the hypermomentum
$\Delta^\alpha{}_\beta:=\delta L_{\rm Max}/\delta
\Gamma_\alpha{}^\beta$ vanishes, $\Delta^\alpha{}_\beta=0$, and the
only external current is the electromagnetic energy-momentum current
$\Sigma_\alpha=\stackrel{\rm Max}{T_\alpha}$. We will only be able to
find non-trivial solutions, if the coupling constants fulfill the
constraint
\begin{equation}
  \gamma =-\frac{8}{3}\frac{\beta}{\beta +6}\;.
          \label{constraint}
\end{equation}
This is the best we can do so far.

It has been shown by Obukhov et al.\ \cite{OVEH} --- for the related
work of Tucker et al., see \cite{TuckerJadwisin} --- how the {\em
  general} solution belonging to Lagrangians of the type (\ref{Ltot1})
can be constructed: One starts with an electrovac metric of general
relativity and the corresponding electromagnetic potential.  We will
take the P\&D metric (\ref{pd}) and its potential (\ref{W}).  Then the
post-Riemannian triplet of (\ref{QQ},\ref{TT}) is constructed which is
patterned after the electromagnetic potential,
\begin{equation}\label{tripp}
Q=k_0\,\omega\,,\qquad\Lambda=k_1\,\omega\,,\qquad T=k_2\,\omega\,,
\end{equation} with the 1-form
\begin{eqnarray}
  {\omega}&=& \frac{1}{\Delta}\left[( N_{\rm e}\,q + N_{\rm g}\,p)\, {
      d \tau} + (N_{\rm g}\,q - N_{\rm e}\,p)\, p \,q \,{ d \sigma}
  \right ]\nonumber\\ &=& \frac{H}{\sqrt{\Delta}}\left( \frac{N_{\rm
        e}\, q }{\sqrt{{\cal Q}}} \;\vartheta^{\hat{0}} + \frac{N_{\rm
        g}\, p}{\sqrt{{\cal P}}} \;\vartheta^{\hat{3}}\right)\,.
\label{Womega}
\end{eqnarray}
In Excalc this input reads:
\begin{verbatim}
%***************************************************************
%       Triplet ansatz for nonmetricity and torsion            *
%***************************************************************
% file triplet.exi, 1998-03-15
% prior input coframe o(a), frame e(a), metric g(a,b)

pform {omega1,qq1,llam1,tt1}=1$

omega1    := (1/delta)*(  (Ne*q+Ng*p)     * d tau 
                        + (Ng*q-Ne*p)*p*q * d sigma); 

qq1   :=  k0 * omega1$
llam1 :=  k1 * omega1$
tt1   :=  k2 * omega1$

clear omega1$
$end$
%***************************************************************
\end{verbatim}
These are the three one-forms which are fundamental for the exact
solution. Now these one-forms have to be assigned to the corresponding
pieces of the torsion $T^\alpha\sim {\tt torsion2(a)}$, the contortion
$K_{\alpha\beta}\sim{\tt contor1(-a,-b)}$, and the nonmetricity
$Q^{\alpha\beta}\sim{\tt nonmet1(a,b)}$:
\begin{verbatim}
%***************************************************************
%                Post-Riemannian assignments                   *
%***************************************************************
% file postriem.exi, 1998-03-15
% prior input o(a), e(a), g(a,b,), qq1, llam1, tt1

pform torsion2(a)=2,{contor1(a,b),nonmet1(a,b)}=1$

torsion2(a)   := (1/3) * o(a) ^ tt1$

nonmet1(a,b)  := (2/9)*( o(b) * (e(a) _| llam1) 
                        +o(a) * (e(b) _| llam1) 
                        -(1/2)*g(a,b)*llam1 ) + qq1*g(a,b)$

contor1(-a,-b) :=  (1/2)*(e(-a) _| torsion2(-b)
                       -  e(-b) _| torsion2(-a)) 
                  -(1/2)*(e(-a) _| (e(-b)_|torsion2(-c)))^o(c)$
$end$
%***************************************************************
\end{verbatim}
The constants $k_0,k_1,k_2$ in front of the triplet can be expressed
in terms of the coupling constants of the dilation-shear Lagrangian:
\begin{equation}\label{k0k1k2}
  k_0=-\frac{24}{\beta+6}\,,\quad k_1=-\frac{36\beta}{\beta+6}\,,
  \qquad k_2=6\,.
\end{equation}
$N_{\rm e}$ and $N_{\rm g}$ are the quasi-electric and the
quasi-magnetic dilation--shear--spin charges, respectively.  Now, in
metric-affine gravity, the polynomials depend also on these
quasi-charges in a fairly trivial, i.e.\ in an additive way,
\begin{eqnarray}
  {\cal P} &:=& \left(b - g_{\rm o}^2 - {\cal G}_{\rm o}^2\right) +
  2np - \epsilon p^2 + 2 m \mu p^3 - \left[\mu^2 \left(b + e_{\rm
        o}^2+{\cal E}_{\rm o}^2 \right) \right] p^4 ,\nonumber\\ {\cal
    Q} &:=& \left(b + e_{\rm o}^2 +{\cal E}_{\rm o}^2\right) - 2mq +
  \epsilon q^2 - 2n\mu q^3 - \left[ \mu^2 \left(b -g_{\rm o}^2-{\cal
        G}_{\rm o}^2 \right) \right] q^4 ,\nonumber \\ \Delta &:=& p^2
  + q^2 ,\nonumber \\ H &:=& 1 - \mu \, p \, q\,.
\label{solution-n}
\end{eqnarray}
The post-Riemannian charges $N_{\rm e}$ and $N_{\rm g}$ are related to
the post-Riemannian pieces ${\cal E}_{\rm o}$ and ${\cal G}_{\rm o}$,
entering the polynomials, according to
\begin{equation}\label{constraint'}
  {\cal E}_{\rm o}=k_0\sqrt{\frac{\alpha}{2}}\,N_{\rm e}\,,\qquad
  {\cal G}_{\rm o} =k_0\sqrt{\frac{\alpha}{2}}\,N_{\rm g}\,,\qquad{\rm
    with}\qquad \alpha >0\,.
\end{equation}

The field equations of metric-affine gravity can be expressed in term
of the irreducible components of curvature, torsion, and nonmetricity
-- and these quantities we have already computed. With a corresponding
computer program we verified the correctness of the exact solution
discussed in this section. The presentation of these programs,
however, we will leave to another occasion.

\section{Discussion}

Effective work in gravity can nowadays only be done if one has the
standard computer algebra methods and packages available. We hope to
have shown that the Reduce-Excalc tool is very efficient in the
context of the Einstein-Maxwell equations as well as in gauge theories
of gravity, like metric-affine gravity, where the spacetime is a
non-Riemannian manifold.

\begin{ack} We are grateful to Christian Heinicke and Yuri Obukhov for 
  interesting discussions and useful remarks. This work was partially
  supported by CONACyT grants 3544--E9311, 3898P--E9608, and by the
  joint Mexican--German project CONACyT--DLR E130--2924, DLR--CONACyT
  6.B0a.6A. Moreover, J.S.\ acknowledges support from the ANUIES--DAAD
  agreement, Kennziffer A/98/04459.
\end{ack}

\centerline{ =============================}

\begin{thebibliography}{9}

\bibitem{TCLASSI} J.E. {\AA}man, J.B.~Fonseca-Neto, M.J.~Rebou\c{c}as,
  M.A.H.~MacCallum: ``TCLASSI: A Computer Algebra Package for Torsion
  Theories of Gravitation''. Preprint (April 1995).

\bibitem{Aman} J.E.  {\AA}man, J.B. Fonseca-Neto, M.A.H. MacCallum,
  M.J.  Reboucas: ``Riemann-Cartan space-times of Godel type''.  Los
  Alamos eprint archive gr-qc/9711064 (1997).

\bibitem{Brans} C.H. Brans: ``Computer Algebra and General
  Relativity''. In Ref.\cite{Fleischer}, pp.\ 183--194.

\bibitem{Debever0} R. Debever, N. Kamran, R.G. McLenaghan: ``A single
  expression for the general solution of Einstein's vacuum and
  electrovac field equations with cosmological constant for Petrov
  type D admitting a nonsingular aligned Maxwell field''.  Phys. Lett.
  {\bf 93A} (1983) 399-402.

\bibitem{Debever} R. Debever, N. Kamran, R.G. McLenaghan: ``Exhaustive
  integration and a single expression for the general solution of the
  type D vacuum and electrovac field equations with cosmological
  constant for a nonsingular aligned Maxwell field''. J. Math. Phys.
  {\bf 25} (1984) 1955-1972.

\bibitem{TuckO} T. Dereli, M. \"Onder, J.  Schray, R.W. Tucker, C.
  Wang: ``Non-Riemannian gravity and the Einstein-Proca system''.
  Class. Quantum Grav. {\bf 13} (1996) L103--L109.

\bibitem{Fleischer} J.~Fleischer, J. Grabmeier, F.W. Hehl, W.
  K\"uchlin, eds.: {\em Computer Algebra in Science and Engineering}
  (World Scientific, Singapore, 1995).

\bibitem{GR14} M. Francaviglia, G. Longhi, L. Lusanna, E. Sorace,
  eds.: {\em Proc.\ of the 14th International Conference on 
      General Relativity and Gravitation (GR14)}. Florence 6-12 August
    1995 (World Scientific, Singapore, 1997).

\bibitem{Alberto} A. Garc\'{\i}a D\'{\i}az: ``Electrovac type D
  solutions with cosmological constant''. J. Math. Phys. {\bf 25}
  (1984) 1951--1954.

\bibitem{blackhole} A. Garc\'{\i}a and A. Mac\'{\i}as: ``Black Holes
  as exact solution of the Einstein--Maxwell equations of Petrov type
  D''.  In: {\em Black Holes: Theory and Observation}. F.W.  Hehl, C.
  Kiefer, R. Metzler, eds. (Springer, Berlin, 1998) in print.

\bibitem{Erice95} F.~Gronwald and F.W.~Hehl: ``On the gauge aspects of
  gravity''. In: \emph{International School of Cosmology and
    Gravitation:} 14\({}^{\rm th}\) Course: Quantum Gravity, held May
  1995 in Erice, Italy.  Proceedings. P.G. Bergmann et al.(eds.)
  (World Scientific, Singapore, 1996) pp.\ 148--198. Los Alamos eprint
  archive gr-qc/9602013.

\bibitem{David} D. Hartley: ``Overview of Computer Algebra in
  Relativity''.  In Ref.\cite{Honnef}, pp.\ 173--191.

\bibitem{Hearn} A.C. Hearn {\em {REDUCE} User's Manual, Version 3.5}
  RAND Publication CP78 (Rev. 10/93) The RAND Corporation, Santa
  Monica, CA 90407-2138, USA (1993).

\bibitem{GR14CA} F.W. Hehl: ``Computer methods in general relativity:
  algebraic computing''. Workshop A.5(ii) of GR14, see
  Ref.\cite{GR14}, pp.\ 469--477.

\bibitem{jadwisin} F.W. Hehl, J. Socorro: ``Gauge theory of gravity:
  electrically charged solutions within the metric-affine framework''.
  Acta Physica Polonica {\bf B}, to be published (1998).  Los Alamos
  eprint archive gr-qc/9803037.

\bibitem{PRs} F.W.\ Hehl, J.D.\ McCrea, E.W.\  Mielke, Y.\ Ne'eman:
  ``Metric-affine gauge theory of gravity: Field equations, Noether
  identities, world spinors, and breaking of dilation invariance''.
  Phys. Rep. {\bf 258} (1995) 1--171.

\bibitem{Honnef} F.W. Hehl, R.A. Puntigam, H. Ruder, eds.: {\em
    Relativity and Scientific Computing -- Computer Algebra, Numerics,
    Visualization} (Springer, Berlin, 1996).

\bibitem{Heinicke} C. Heinicke: ``The Plebanski-Demianski metric''.
  Unpublished manuscript of a mini-research project. University of
  Cologne (March 1998).

\bibitem{Klioner} S.A.\ Klioner: ``EinS: A Mathematica Package for
  Tensorial Calculations in Astronomical Applications of Relativitstic
  Gravity Theories''. Abstract submitted to GR14. Email: {\tt
    klioner@ipa.rssi.ru}.

\bibitem{W+A} W. Kopczy\'nski, A. Trautman: {\it Spacetime and
    Gravitation}, translated from the Polish (Wiley, Chichester, and
  PWN -- Polish Scientific Publishers, Warszawa, 1992).

\bibitem{ma} A. Mac\'{\i}as, E. W. Mielke and J. Socorro: ``Solitonic
  monopole solution in metric-affine gauge theory carrying Weyl
  charges''. Class. Quantum Grav. {\bf 15} (1998) 445-452.

\bibitem{Dermott} J.D. McCrea: ``REDUCE in General Relativity and
  Poincar\'e Gauge Theory''. In Ref.\cite{rio}, pp.\ 173--263. See
  the library {\tt ftp://euclid.maths.qmw.ac.uk/pub /grlib}.

\bibitem{McL} R.G. McLenaghan: ``MAPLE applications to general
  relativity''. In Ref.\cite{rio}, pp. 265--354.

\bibitem{OVEH} Yu.N. Obukhov, E.J. Vlachynsky, W. Esser, F.W.  Hehl:
  ``Effective Einstein theory from metric-affine gravity models via
  irreducible decompositions''. Phys. Rev. {\bf D56} (1997)
  7769--7778.

\bibitem{OVETH} Yu.N. Obukhov, E.J. Vlachynsky, W. Esser, R.
  Tresguerres, F.W. Hehl: ``An exact solution of the metric-affine
  gauge theory with dilation, shear, and spin charges''.  Phys. Lett.
  {\bf A220} (1996) 1--9.

\bibitem{Parker} L. Parker, S.M. Christensen: {\em MathTensor: A
  System for Doing Tensor Analysis by Computer} (Addison-Wesley,
  Redwood City, 1994).

\bibitem{pd} J.F. Plebanski and M. Demianski: ``Rotating, charged, and
  uniformly accelerating mass in general relativity''. Ann. Phys.
  (N.Y.)  {\bf 98} (1976) 98--127.

\bibitem{Pollney} D. Pollney, P. Musgrave, K. Santosuosso, K. Lake:
  ``Algorithms for computer algebra calculations in space-time: I.
  The calculation of curvature''. Class. Quant. Grav. {\bf 13} (1996)
  2289-2310. Los Alamos eprint archive gr-qc/9601036.

\bibitem{PLH97}R.A. Puntigam, C. L\"ammerzahl and F.W. Hehl:
  ``Maxwell's theory on a post--Riemannian spacetime and the
  equivalence principle''. Class. Quant.  Grav. {\bf 14} (1997)
  1347--1356.

\bibitem{PuntigamSH} R.A. Puntigam, E. Schr{\"u}fer, F.W. Hehl: ``The
  use of computer algebra in {M}axwell's theory''. In
  Ref.\cite{Fleischer}, pp.\ 195-211

\bibitem{rio} M.J.~Rebou\c{c}as, W.L.~Roque, eds.: {\em Algebraic
    Computing in General Relativity (Lecture Notes from the First
    Brazilian School on Computer Algebra, vol.\ 2)} (Oxford University
  Press, Oxford, 1994).

\bibitem{Schruefer} E. Schr\"ufer: {\em EXCALC: A System for Doing
    Calculations in the Calculus of Modern Differential Geometry}.
  GMD-SCAI, D-53757 St.Augustin, Germany (1994).

\bibitem{EXCALC} E. Schr\"ufer, F.W. Hehl, J.D. McCrea: ``Exterior
  Calculus on the Computer: The REDUCE--package EXCALC Applied to
  General Relativity and the Poincar\'e Gauge Theory''. Gen.\ Rel.\ 
  Grav.\ {\bf 19} (1987) 197--218.

\bibitem{Harald} H.H.~Soleng: ``The Mathematica Packages CARTAN and
  MathTensor for Tensor Analysis''. In Ref.\cite{Honnef}, pp.\ 
  210--230.

\bibitem{Stauffer} D. Stauffer, F.W. Hehl, N. Ito, V. Winkelmann, J.G.
  Zabolitzky: {\it Computer Simulation and Computer Algebra --
    Lectures for Beginners.} 3rd ed.\ (Springer, Berlin, 1993).

\bibitem{Irina} S.I. Tertychniy and I.G. Obukhova: {\em GRG$_{\rm
      EC}$: Computer Algebra System for Applications in Gravitation
    Theory.} {\sl SIGSAM Bulletin} {\bf 31}, n.\ 1, issue 119 (1997)
  6-14.

\bibitem{Tres1} R. Tresguerres: ``Exact vacuum solutions of
  4--dimensional metric--affine gauge theories of gravitation''.  Z.
  Phys. {\bf C65} (1995) 347--354.

\bibitem{Tres2} R. Tresguerres: ``Exact static vacuum solution of
  four--dimensional metric--affine gravity with nontrivial torsion''.
  Phys. Lett. {\bf A200} (1995) 405--410.

\bibitem{Tsantilis} E. Tsantilis, R.A. Puntigam, F.W.\ Hehl: ``A
  quadratic curvature Lagrangian of Paw{\l}owski and R\c{a}czka: A
  finger exercise with MathTensor''. In Ref.\cite{Honnef}, pp.\ 
  231--240.

\bibitem{TuckerJadwisin} R. Tucker: Talk given at the Workshop on
    {\em Gauge Theories of Gravitation,} Jadwisin, Poland, 4 to
    10-Sept-1997.
     
\bibitem{TuckerWang} R.W.  Tucker and C.  Wang: ``Black holes with
    Weyl charge and non-Riemannian waves''. Class. Quantum Grav. {\bf
      12} (1995) 2587--2605.

\bibitem{VTOH} E.J. Vlachynsky, R. Tresguerres, Yu.N. Obukhov, F.W.
    Hehl: ``An axially symmetric solution of metric-affine gravity''.
    Class. Quantum Grav. {\bf 13} (1996) 3253--3259.

\bibitem{Wolf} T.~Wolf: ``The Program CRACK for Solving PDEs in
    General Relativity''. In Ref.\cite{Honnef}, pp.\ 241--258.

\bibitem{GRG} V.V.~Zhytnikov: {\em GRG: Computer Algebra System for
      Differential Geometry, Gravitation and Field Theory. Version
      3.1} (Moscow, 1992) GRG3.1 is available via anonymous ftp from
    {\tt ftp.maths.qmw.ac.uk} in the directory {\tt pub/grg3.1}.
\end{thebibliography}
\end{document}